\documentclass[1p,authoryear]{elsarticle}

\usepackage{amsfonts,amssymb,amsmath}

\newcommand{\at}{{\char '100}}
\newcommand{\EXP}[1]{\mathrm{e}^{#1}} 
\newcommand{\DEF}{\overset{\mathrm{def}}{=}}
\newcommand{\DEFt}{\smash{\overset{\text{\tiny def}}{=}}}
\newcommand{\imat}{{\mathrm{i}}} 

\newcommand{\im}{\operatorname{Im}}
\newcommand{\re}{\operatorname{Re}}

\newcommand{\ket}[1]{|#1\rangle} 
\newcommand{\kete}[1]{|\kern.3ex#1\kern.3ex\rangle}
\newcommand{\bra}[1]{\langle #1 |}
\newcommand{\brae}[1]{\langle\kern.3ex #1 \kern.3ex|} 
\def\braket#1#2{\mathinner{\langle{#1}|{#2}\rangle}}
\newcommand{\T}[1]{#1^{\scriptscriptstyle \mathrm{T}}} 
\newcommand{\oldstyle}[1]{{#1}}
\newcommand{\nompropre}[1]{{#1}}

\begin{document}

\title{An alternative proof of Wigner theorem on quantum transformations based on elementary complex analysis}
\author{Amaury Mouchet}
\ead{mouchet\at lmpt.univ-tours.fr}
\address{Laboratoire de Math\'ematiques
  et de Physique Th\'eorique\\
Universit\'e Fran\c{c}ois Rabelais de Tours --- \textsc{\textsc{cnrs (umr 7350)}}\\
F\'ed\'eration Denis Poisson\\
 Parc de Grandmont 37200\\
  Tours,  France.}

\begin{abstract}
According to Wigner theorem, transformations of quantum states which preserve
the probabilities are either unitary or antiunitary. 
This short communication presents an elementary proof of this theorem
that significantly departs from the numerous 
ones already existing in the literature. The main line of the argument remains
valid even in quantum field theory where Hilbert spaces are non-separable.
\end{abstract}
\begin{keyword} Wigner theorem \sep symmetry \sep unitary representations

\textit{Highlights:} A new elementary proof of Wigner theorem  
on quantum transformations is presented. It relies on substantially different
hypothesis from the previous proofs. It allows a straightforward generalisation
to non-separable Hilbert spaces.

\textsc{pacs:} 11.30.-j  
03.65.Ta 	
03.70.+k 	
03.65.-w 	

\end{keyword}

\maketitle 

Corresponding author : Amaury Mouchet, mouchet\at lmpt.univ-tours.fr, 
phone:  +33-2.47.36.73.70., fax: 33-2.47.36.69.56.  

\section{Preliminaries}

Wigner theorem is a cornerstone of theoretical physics since it encapsulates
all the linear structure of quantum transformations, among which
the evolution of quantum systems (aside from the measurement process).
More precisely, given a Hilbert space~$\mathcal{H}$ endowed with a Hermitian
product~$\braket{\chi}{\psi}$

Wigner, in the early 30's \citep[Appendix to chap.~20, pp~233--236, for the updated English translation]{Wigner59a}, acknowledged that any transformation~$T:\ket{\psi}\mapsto \T{\ket{\psi}}=T\big(\ket{\psi}\big)$
such that 
\begin{equation}\label{eq:conservationproba}
  \forall \big(\ket{\chi},\ket{\psi}\big)
\in\mathcal{H}^2, \qquad \big|\T{\;}\!\!\T{\braket{\chi}{\psi}}\big|
   =\big|\braket{\chi}{\psi}\big|
\end{equation}  
is either 

(a) linear   $T\big(c_1\ket{\psi_1}+c_2\ket{\psi_2}\big)
                      =\big(c_1 \T{\ket{\psi_1}}+c_2\T{\ket{\psi_2}}\big)$ and unitary 
                    $T^{-1}=T^*$;

\noindent or 

(b) antilinear    $T\big(c_1\ket{\psi_1}+c_2\ket{\psi_2}\big)
                      =c^*_1 \T{\ket{\psi_1}}+c^*_2\T{\ket{\psi_2}}$ and unitary
                    $T^{-1}=T^*$ (such a map is also called antiunitary).

We shall systematically use the usual Dirac bra-ket notation and, in
the above definitions, $(c_1,c_2)$ stands for any pair of complex
coefficients, $\big(\ket{\psi_1},\ket{\psi_2}\big)$ is any pair of elements
of $\mathcal{H}$. The Hermitian conjugate will be denoted by $(\ )^*$
and therefore, if $z$ is just one complex number, $z^*$ stands for its
complex conjugate.

Since Wigner's original work that pertained to the representation of
the  rotation group, many proofs and generalizations have been
proposed whose levels of rigour are not necessarily correlated to
their length but, rather, vary depending on the concern of their
author. See for
  example  \citep{Simon+08a} and  references therein to which I shall add
\citep[\S~1.3.2]{Fonda/Ghirardi70a} and the concise and elegant proof
given by \citep[\S~XV-2]{Messiah59a}.  The present work is further added to this list because it
appears to be almost a back-of-the-envelope presentation while keeping
a level of rigour that is acceptable by physicists (hopefully, little
additionnal work on the main key ideas should meet to the requirements
of mathematicians as well).  Moreover, the majority of the previous
proofs, if not all of them, can hardly be transposed to non-separable
Hilbert spaces, that is to spaces where a countable orthonormal basis
does not exist. Yet, in quantum field theory, such non-separable
Hilbert spaces are unavoidable: any continuous canonical
transformation or rearrangement of the infinitely many degrees of
freedom --- that physically describe a renormalisation
of the bare particles into the dressed ones like, for instance, when a condensation occurs ---
requires that the Hilbert space is made of a 
continuous family of orthogonal Fock
spaces and one cannot content one self with the unique Fock space that
represents the physical (dressed) particles; see for example 
\citep[\S1.1]{Emch72a} or \citep[chap.~3]{Umezawa93a}. 

The proof presented in the second part of the present article will at first 
make use of a (at most) 
countable basis $\big(\ket{\varphi_\nu}\big)_{\nu\in I}$ 
in $\mathcal{H}$,
\begin{equation}\label{eq:orthormalization}
  \braket{\varphi_\nu}{\varphi_\mu}=\delta_{\nu\mu}\;,
\end{equation}
but the discreteness of the set~$I$  will be just a matter of
convenience. The transposition to a continuous (multi-)index is
straightforward (we will not use any induction arguments that would
prevent such generalization) and physicists are quite used to it.  For
instance, with an appropriate choice of normalization, the Kronecker
symbol in \eqref{eq:orthormalization} is replaced by a Dirac
distribution, the discrete sequences~$z=(z_\nu)_{\nu\in I}$
labelled by~$\nu$ become regular 
functions~$z:\nu\mapsto z(\nu)$, functions of~$z$ become functionals, matrices are turned into operators, etc.

Unlike in many proofs,  we
 shall not suppose a priori that~$T$ is bijective. The only
additional property will be that~$T$ is differentiable twice 
(just once may be sufficient but 
we will not try no minimize the requirements on the regularity of the transformation). 
Let us just mention that
such smoothness is a reasonable supposition based on physical grounds.
Except during the measuring process, when the number of 
degrees of freedom involved in the interaction of the system with 
a measuring device becomes infinite --- and we know that the superposition
principle as well as condition~\eqref{eq:conservationproba} 
are lost by the transformation on the states 
induced by the measuring process ---, we have never observed
any discontinuity nor singularity 
with respect to the quantum state in a transformation. 

To see how this differentiable hypothesis simplifies considerably the proof 
of Wigner theorem by offering a simple strategy, let us see how this works
 in the
Euclidean case where the analogous of the Wigner theorem is known as
 Mazur-Ulam theorem\citep[this first version, concerning more generally isometries in real normed vector spaces, was published the following
year after Wigner's]{MazurUlam32a}.  
Let us suppose that~$T$ is a differentiable application in an Euclidean space~$\mathcal{E}$
preserving the scalar product ``$\cdot$'' i.e. such that 
\begin{equation}\label{eq:isometry}
    \forall \big(u,v\big)
\in\mathcal{E}^2, \qquad T(u)\cdot T(v)=u\cdot v\;.
\end{equation}  

Then, differentiating~\eqref{eq:isometry} with respect to~$u$ and~$v$ leads to
\begin{equation}\label{eq:realTprime}
   \forall \big(u,v\big)
\in\mathcal{E}^2, \qquad \big(T'(u)\big)^{\mathrm{t}}\; T'(v)=1
\end{equation}
(``${\ }^{\mathrm{t}}$\;'' denotes the transposition and ``${\ }'$ '' the derivative) which shows immediately that
the Jacobian matrix~$T'(v)$ is invertible, independent 
of~$v$ and indeed orthogonal. Since moreover $T(0)=0$ (the scalar product is non degenerate),
 we have proven that $T$~is necessary a linear
orthogonal transformation (it is obviously a sufficient condition).

Under the alternative hypothesis of being surjective,  
the Mazur-Ulam theorem is also immediate since,
then, the orthonormal basis~$(e_\nu)_{\nu\in I}$ is mapped to an orthonormal basis~$(\T{e_\nu})_{\nu\in I}$
on which, for any~$v$ in~$\mathcal{E}$, we can expand $\T{v}$ and then, 
\begin{equation}\label{eq:Tv_mazurUlam}
 \T{v}=\sum_\nu (\T{e_\nu}\cdot \T{v})\;\T{e_\nu}=\sum_\nu (e^{}_\nu\cdot v)\;\T{e_\nu}\;.
\end{equation} 
If there were a non-null vector~$w$ in~$\mathcal{E}$ that were orthogonal to the subspace
spanned by~$(\T{e_\nu})_{\nu\in I}$, then its preimage $T^{-1}(w)$, whose existence is guaranteed by the surjectivity hypothesis,
would be the null vector, being a vector orthogonal to any~$e_\nu$. This is in contradiction with~$T(0)=0$. 
Relation~\eqref{eq:Tv_mazurUlam} can be written as
\begin{equation}
  T\left(\sum_\nu (e_\nu\cdot v)\;e_\nu\right)=\sum_\nu (e_\nu\cdot v)\;T({e_\nu})
\end{equation}
which expresses the linearity of~$T$ whose orthogonality follows.
However, in the Hermitian case, the possible phase factors that may appear in the relation between
$\T{\rule{0cm}{1.7ex}}\!\!\T{\braket{\varphi_\nu}{\psi}}$ and~$\braket{\varphi_\nu}{\psi}$  make
 the proofs following the second line of thought much less straightforward. 

\section{The core of the proof}

\paragraph{The invariance property  formulated in terms of complex analysis} 
We shall take advantage of the isomorphism that allows us to identify the Hilbert space~$\mathcal{H}$ to
the set of sequences of complex numbers~$z=(z_\nu)_{\nu\in I}$. More precisely, once given an orthonomal basis
$\big(\ket{\varphi_\nu}\big)_{\nu\in I}$, we can map any $\ket{\psi}$ in~$\mathcal{H}$ to~$z$ 
with~$z_\nu=\braket{\varphi_\nu}{\psi}$. The Hermitian product in~$\mathcal{H}$ reads
\begin{equation}
     \braket{w}{z}\DEF w^*\; z=\sum_\nu w_\nu^*\;z^{}_\nu=\braket{\chi}{\psi}
\end{equation} 
with $z=\big(\braket{\varphi_\nu}{\psi}\big)_{\nu\in I}$ and
$w=\big(\braket{\varphi_\nu}{\chi}\big)_{\nu\in I}$. Any application~$T$ in~$\mathcal{H}$ can be seen as a either a
function~$\mathcal{T}(x,y)$ of the sequence of the real part~$x=\re z$ and imaginary part~$y=\im z$ of the components of the state, or
rather as a function of two \textit{independent} complex sequences~$z$ and~$\bar{z}$ 
\begin{equation}\label{def:Tzzbar}
  T(z,\bar{z})\DEF\mathcal{T}\big((\bar{z}+z)/2,\imat(\bar{z}-z)/2)
\end{equation}
evaluated at~$\bar{z}=z^*$. 
When~$\mathcal{T}$ can be differentiated, we can define the derivatives with respect to the complex variables
by \citep[\S~12.1, for instance]{Dubrovine84a+}
\begin{subequations}\label{def:ddbar}
\begin{eqnarray}
       \partial_z T &\DEF& \frac{1}{2}\left(\partial_x-\imat\partial_y\right)\mathcal{T}\;;\\
       \partial_{\bar{z}} T &\DEF& \frac{1}{2}\left(\partial_x+\imat\partial_y\right)\mathcal{T}
       \label{subeq:dbarf}\;.
\end{eqnarray}
\end{subequations}
In particular, $T$ is analytic if and only if~$\partial_{\bar{z}} T=0$.
From the differential of~$T$, we get 
\begin{equation}\label{eq:diffzetoile}
       \partial_{z^*}\big(T(z,z^*)\big)=(\partial_{\bar{z}}T)(z,z^*)\;. 
\end{equation}
In the following we will drop the distinction 
between~$z^*$ and the variable~$\bar{z}$ that may vary independently of~$z$ because
the continuation of any application~$T(z,\bar{z})$  defined by~\eqref{def:Tzzbar}
in the domain where~$\bar{z}\neq z^*$ is unique. 

In this language, 
the invariance condition \eqref{eq:conservationproba} can be reformulated as follows,
\begin{equation}\label{eq:modulusTstarT}
        \forall (w,z)\in\mathcal{H}^2, \qquad \big|\big(T(w,w^*)\big)^*\; T(z,z^*)\big|=|w^*\; z|\;.
\end{equation}
Therefore we have two possibilities : there exists a real function~$\theta$ of the complex 
variables~$(w,w^*,z,z^*)$ such that
\begin{subequations}
\label{eq:wignerT}
\begin{align}
&\text{(a) either }& \big(T(w,w^*)\big)^*\; T(z,z^*)=&\ \EXP{\imat\theta(w,w^*,z,z^*)}w^*\; z\;;&
\label{eq:wignerTa}\\
&\text{(b) or}& \big(T(z,z^*)\big)^*\;T(w,w^*)=&\ \EXP{-\imat\theta(w,w^*,z,z^*)}w^*\; z\;.&
\label{eq:wignerTb}
\end{align}
\end{subequations}
To understand where these two conditions come from, first divide the both sides of the equality
in~\eqref{eq:modulusTstarT} by~$|w^*\; z|$ when non vanishing (then eventually
include these cases by continuity of the transformation~$T$). Thus, we are led to
an equation of the form~$|Z|=1$ where~$Z$ is a complex number. Then, 
the argument~$\eta$ of~$Z$ can take the two solutions $\pm\arccos(\re{Z})$ 
in $[-\pi,\pi]$ of the equation~$\cos\eta=\re{Z}$. The two options~(a) and~(b)
correspond to the two forms $\EXP{\pm\imat\arccos(\re{Z})}$ that is 
$Z=\EXP{\imat\arccos(\re{Z})}$ or $Z^*=\EXP{\imat\arccos(\re{Z})}$.

Permuting~$(w,w^*)$ and~$(z,z^*)$, then by conjugation we have necessarily
\begin{equation}\label{eq:thetaantisym}
        \theta(z,z^*,w,w^*)=-\theta(w,w^*,z,z^*)
\end{equation}
which incidentally shows that~$\theta(0,0,0,0)=0$.  

\paragraph{Adjusting the phase} The first step of our proof is to
  redefine the phase of the transformed states by considering
\begin{equation}
        \widetilde{T}(z,z^*)=\EXP{\imat\alpha(z,z^*)}T(z,z^*)
\end{equation}
where~$\alpha$ is a real function. Conditions~\eqref{eq:wignerT} can be written
\begin{subequations}
\begin{align}
&\text{(a) either }& \big(\widetilde{T}(w,w^*)\big)^*\;\widetilde{T}(z,z^*)
=&\ \EXP{\imat\tilde{\theta}(w,w^*,z,z^*)}w^*\; z\;;&
\\
&\text{(b) or}& \big(\widetilde{T}(z,z^*)\big)^*\;\widetilde{T}(w,w^*)
=&\ \EXP{-\imat\tilde{\theta}(w,w^*,z,z^*)}w^*\; z\;.&
\end{align}
\end{subequations}
with
\begin{equation}
    \tilde{\theta}(w,w^*,z,z^*)=\theta(w,w^*,z,z^*)+\alpha(z,z^*)-\alpha(w,w^*)\;.
\end{equation} 
The latter identity can be used to cancel~$\tilde{\theta}(0,0,z,z^*)$ with the choice
\begin{equation}
     \alpha(z,z^*)=-\theta(0,0,z,z^*)=\theta(z,z^*,0,0)\;.
\end{equation}
 Therefore
 without loss of generality, in the following we can consider that for all~$z$
\begin{equation}\label{eq:thetawzerozzetoile}
     \theta(0,0,z,z^*)=0\;.
\end{equation}

\paragraph{Differentiation} The second step of the proof is to differentiate the conditions
written in~\eqref{eq:wignerT} with respect to~$w^*$ and~$z$. From
\eqref{eq:wignerTa}, using
$\partial_{w^*}\big(T(w,w^*)\big)^*=\big(\partial_{w}T(w,w^*)\big)^*$,
we get
\begin{multline}\label{eq:diffTwwetoileTzzetoile}
      \big(\partial_{w}T(w,w^*)\big)^* \partial_{z}T(z,z^*)=\EXP{\imat\theta(w,w^*,z,z^*)}\Big\{ 1+\\ 
+ \imat w^* \partial_{w^*}\theta(w,w^*,z,z^*)  + \imat \partial_{z}\theta(w,w^*,z,z^*) z\\
+\big[\imat \partial^2_{z,w^*}\theta(w,w^*,z,z^*)\\
 -\partial^{}_{z}\theta(w,w^*,z,z^*)\;\partial^{}_{w^*}\theta(w,w^*,z,z^*)\big] w^*\; z\Big\}.
\end{multline}
Because of this last computation, and in particular because of the
presence of the Hessian of~$\theta$, we assume that the transformation
is differentiable twice. When evaluated for~$(w,w^*)=(0,0)$, we get
\begin{equation}\label{eq:complexTprime_a}
        \big(\partial_{z}T(0,0)\big)^*\; \partial_{z}T(z,z^*)=1
\end{equation} 
since~\eqref{eq:thetawzerozzetoile} implies  $\partial_{z}\theta(0,0,z,z^*)=0$.
Then $\big(\partial_{z}T(z,z^*)\big)^{-1}=\big(\partial_{z}T(0,0)\big)^*$ appears to be 
independent of $(z,z^*)$ and unitary. Moreover, because~$T(0,0)=0$, we have, for all~$z$,
\begin{equation}\label{eq:Tzzetoilezerozeroz_a}
        T(z,z^*)=\partial_{z}T(0,0)\; z\;
\end{equation}
and therefore $z\mapsto\T{z}= T(z,z^*)$ is linear. The linear operator~$\hat{U}$ defined by its matrix 
elements~$\bra{\phi_\nu}\hat{U}\ket{\phi_{\nu'}}\DEFt\left(\partial_{z}T(0,0)\right)_{\nu\nu'}$ is such that 
$\hat{U}^{-1}=\hat{U}^*$ and the expression~\eqref{eq:Tzzetoilezerozeroz_a} can be written as
$\T{\braket{\varphi_\nu}{\psi}}
=\sum_{\nu'}\bra{\varphi_\nu}\hat{U}\ket{\varphi_{\nu'}}\braket{\varphi_{\nu'}}{\psi}
=\bra{\varphi_\nu}\hat{U}\ket{\psi}$,
hence $\T{\ket{\psi}}=\hat{U}\ket{\psi}$ which is the first option offered by Wigner theorem.

The second option comes when differentiating~\eqref{eq:wignerTb} which
just modifies the left-hand side of~\eqref{eq:diffTwwetoileTzzetoile}
and the irrelevant sign in front of~$\theta$ in the right-hand
side. Then we obtain
\begin{equation}\label{eq:complexTprime_b}
         \big(\partial_{z^*}T(0,0)\big)^*\;\partial_{z^*}T(z,z^*)=1
\end{equation} 
and eventually 
\begin{equation}\label{eq:Tzzetoilezerozeroz_b}
        T(z,z^*)=\partial_{z^*}T(0,0)\; z^*\;.
\end{equation} 
The  transformation $z\mapsto\T{z}= T(z,z^*)$ is antilinear. An antilinear operator~$\hat{A}$ can be defined from its matrix elements $\bra{\phi_\nu}\big(\hat{A}\ket{\phi_{\nu'}}\big)
\DEFt\left(\partial_{z^*}T(0,0)\right)_{\nu\nu'}$ and we have\linebreak $\T{\braket{\phi_\nu}{\psi}}
=\sum_{\nu'}\bra{\phi_\nu}\left(\hat{A}\ket{\phi_{\nu'}}\right)\braket{\phi_{\nu'}}{\psi}^*
=\bra{\phi_\nu}
\left(\hat{A}\sum_{\nu'}\ket{\phi_{\nu'}}\braket{\phi_{\nu'}}{\psi}\right)= \bra{\phi_\nu}\left(\hat{A}\ket{\psi}\right)$
that is $\T{\ket{\psi}}=\hat{A}\ket{\psi}$. The relation $\big(\partial_{z^*}T(0,0)\big)^{-1}=\big(\partial_{z^*}T(0,0)\big)^*$ reads~$\hat{A}^{-1}=\hat{A}^*$. Therefore case~(b) implies the second alternative of Wigner theorem.

\paragraph{Summary} Looking back the proof over one'shoulder,  once keeping in mind the 
simple line of thought used in the real case --- the double
differentiation of~\eqref{eq:isometry} that leads
to~\eqref{eq:realTprime}---, the backbone of the argument in the
Hilbert case do not appear much more complicated and can be captured in
the following few lines : from the key equations~\eqref{eq:wignerT},
it is straightforward by a double differentiation to
obtain~\eqref{eq:diffTwwetoileTzzetoile}. Then having redefined the
phases of the transformed states in order to get
\eqref{eq:thetawzerozzetoile}, we are immediately led
to~\eqref{eq:complexTprime_a} or~\eqref{eq:complexTprime_b} then to
\eqref{eq:Tzzetoilezerozeroz_a} or \eqref{eq:Tzzetoilezerozeroz_b}
respectively.

\section*{Acknowledgements}

It is a pleasure to thanks Xavier Bekaert for the substantial improvements
he suggested after his careful reading of the first version of this note.


\end{document}